\documentclass{PoS}
\usepackage{amssymb}

\newcommand{\be}{\begin{equation}}
\newcommand{\bea}{\begin{eqnarray}}
\newcommand{\ee}{\end{equation}}
\newcommand{\eea}{\end{eqnarray}}
\renewcommand{\b}{B^0}
\newcommand{\bb}{\bar {B^0}}

\newcommand{\eq}[1]{Eq.\,(\ref{#1})}

\newcommand{\parity}{{\cal P}}
\renewcommand{\and}{\mbox{ and }}

\newcommand{\Dt}{\Delta t}

\newcommand{\w}{\omega}
\newcommand{\Dm}{\Delta m}
\newcommand{\DG}{\Delta \Gamma}
\newcommand{\G}{\Gamma}

\newcommand{\re}{\frac{Re(\epsilon)}{1+|\epsilon|^2}}

\title{The demise of flavour tagging and its $\Dt$-dependence}

\ShortTitle{The demise of flavour tagging and its $\Dt$-dependence}

\author{\speaker{Ezequiel \'Alvarez} \\
        Departamento de Fisica Teorica and IFIC, Universidad de Valencia, CSIC, Spain\\
        E-mail: \email{Ezequiel.Alvarez@uv.es}}

\author{Jose Bernab\'eu\\
        Departamento de Fisica Teorica and IFIC, Universidad de Valencia, CSIC, Spain\\
        }
\author{Miguel Nebot\\
        Centro de F\1sica Te\'orica de Part\1culas, IST, Portugal\\
        }

\abstract{In this work we discuss how the loss of particle-antiparticle identity due to CPT violation affects the observables in the decay of two neutral EPR-correlated $B$-mesons.   We study this possible new effect in the context of equal-sign flavour specific decays and we find a considerable modification in the $\Delta t$-dependence of the equal-sign dilepton charge asymmetry, $A_{sl}$.  Although the more important changes occur right outside the $\Dt$ region that has been up-to-the-date effectively explored, we show that a deeper experimental research may be able to distinguish a possible existence of this new CPT violating parameter, $\w$.  In any case, using the available data for $A_{sl}$, we put the first limits on $\w$.
}

\FullConference{International Europhysics Conference on High Energy Physics\\
		 July 21st - 27th 2005\\
		 Lisboa, Portugal}

\begin{document}

\section{Introduction}
We analyze a novel kind of CPT violation which occurs in the B-factories through the loss of indistingishability of particle-antiparticle \cite{prl}.  

In the usual formulations of {\it entangled} meson states in the B-factories, one imposes the requirement of Bose statistics for the state $\b\bb$, which implies that
the physical system must be $symmetric$ under the combined operation
of charge conjugation ($C$) and permutation of the spatial
coordinates ($\parity$), $C\parity$.  If CPT is violated then $\b$ cannot be considered indistinguishable to $\bb$ and therefore the previous reasoning does not hold any more.  In fact, if the $C\parity=+$ requirement of Bose statistic is relaxed, we can re-write the initial state as
\bea
 |\psi(0)\rangle = \frac{1}{\sqrt{2(1+|\w|^2)}} \times \left\{ | \b (+\vec k),\bb(-\vec k)\rangle - |\bb(+\vec k), \b(-\vec k)\rangle + \right. \nonumber \\
 \left.\w \left( | \b (+\vec k),\bb(-\vec k)\rangle + |\bb(+\vec k), \b(-\vec k)\rangle \right)
 \right\},
 \label{i}
 \eea
where the $\vec k$ vector is along the direction of the momenta of the
mesons in the center of mass system.  Here $\w=|\w| e^{i\Omega}$ is a complex CPT violating parameter, associated with the non-indistinguishable particle nature of the neutral meson
and antimeson states, which parameterizes the loss of Bose
symmetry.  (It is worth noticing that this new parameter, $\w$, it is unrelated to the violation of CPT in the time evolution of the B mixing.)  Observe the symmetry in \eq{i} in which a change in the sign of $\w$ is equivalent to the exchange of the particles $\b
\leftrightarrow \bb$.  Moreover, once defined through \eq{i} the
modulus and phase of $\w$ have physical meaning which could be in
principle measured. 

We interpret \eq{i} as a description of a state of two {\it
distinguishable} particles at first-order in perturbation theory,
written in terms of correlations of the zeroth-order one-particle
states.  As it is easily seen, the probabilities for the two states
connected by a permutation are different due to the presence of
$\w$.  Therefore, in order to give physical meaning to $\w$ we {\it
define}, in \eq{i}, $+\vec k$ as the direction of the {\it first}
decay, and $-\vec k$ as the direction of the second
decay.  Of course, when both decays are simultaneous there is no way
to distinguish the particles; although the effect of $\w$ is still
present \cite{plb}.

It is clear that the modification of the initial state, \eq{i}, will
introduce modifications in all the observables of the $B$ factories,
since it is the departure point of every analysis.

\section{The demise of flavour tagging and the $\Dt$-dependence in the $A_{sl}$ asymmetry}
As it may be seen in \eq{i}, setting $\w\neq 0$ produces a loss in the definite antisymmetry of the state.  This antisymmetry, when $\w=0$, forbids the system to have the same decay
at the same time on both sides,
 \bea
 \langle X,X |U(t)\otimes U(t)|\psi(0)\rangle_{\w=0} = 0,
 \eea
since a permutation of the particles introduces a minus sign in
$|\psi(0)\rangle_{\w=0}$, whereas the rest remains the same.  If, on
the contrary, we have $\w\neq0$, then there is nothing
to be said and the same decay on both sides at the same time it is,
in principle, allowed.  In fact, the time evolution of the initial state gives a projection on the usually {\it forbidden} states $|\b\b\rangle$ and $|\bb\bb\rangle$, proportional to $\w$.  Due to these states, a first flavour-specific $\b$ decay filters at that time the wave function of the meson on the other side to be $\sim {\cal O}(1) |\bb\rangle + {\cal O}(\w)
|\b\rangle$, and vice-verse.  Therefore, the probability of having
the same flavour specific decay {\it at the same time on both sides} goes
as
 \be
 I(\ell^\pm, \ell^\pm, \Dt=0) \sim |\w|^2 .
 \label{demise}
 \ee
This constitutes the demise of flavour tagging \cite{plb}, which introduces conceptual changes in the analysis, but its observability goes as $|\w|^2$.

If we study the time dependence of the intensities in \eq{demise} then the interference terms will give rise to linear correction in $\w$ \cite{next}.  When analyzing the equal-sign dilepton events, $I(\ell^\pm,\ell^\pm,\Dt)$, we find a linear-in-$\w$ {\it opposite} behaviour of the intensities at short $\Dt$'s depending on the sign of the leptons, pointing then to their asymmetry as a good observable where to detect traces of the $\w$-effect.  	

\begin{figure}
 \framebox[\textwidth]{
 \begin{minipage}[c]{1\columnwidth}
   \begin{minipage}[c]{.33\columnwidth}
    \begin{center}
    \includegraphics[width=.94\columnwidth,height=.94\columnwidth]{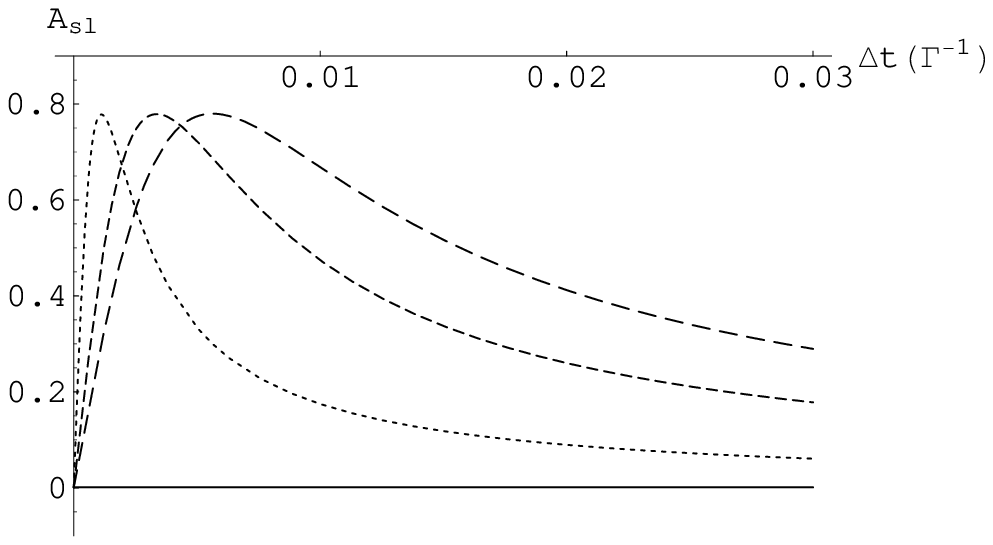}
    \newline
    (a)
    \end{center}
   \end{minipage}
   \begin{minipage}[c]{.33\columnwidth}
    \begin{center}
    \includegraphics[width=.94\columnwidth,height=.94\columnwidth]{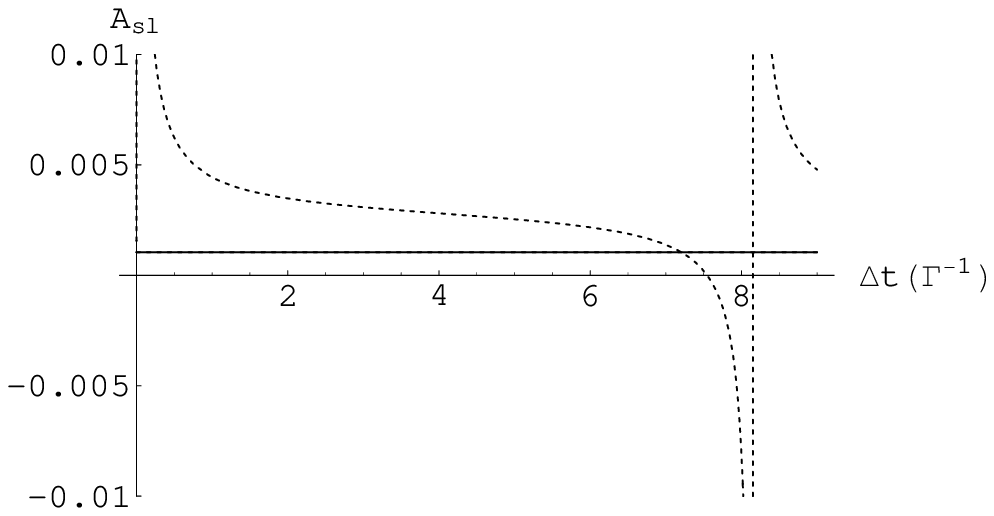}
    \newline 
    (b)
    \end{center}
   \end{minipage}
   \begin{minipage}[c]{.33\columnwidth}
    \begin{center}
    \includegraphics[width=.94\columnwidth,height=.94\columnwidth]{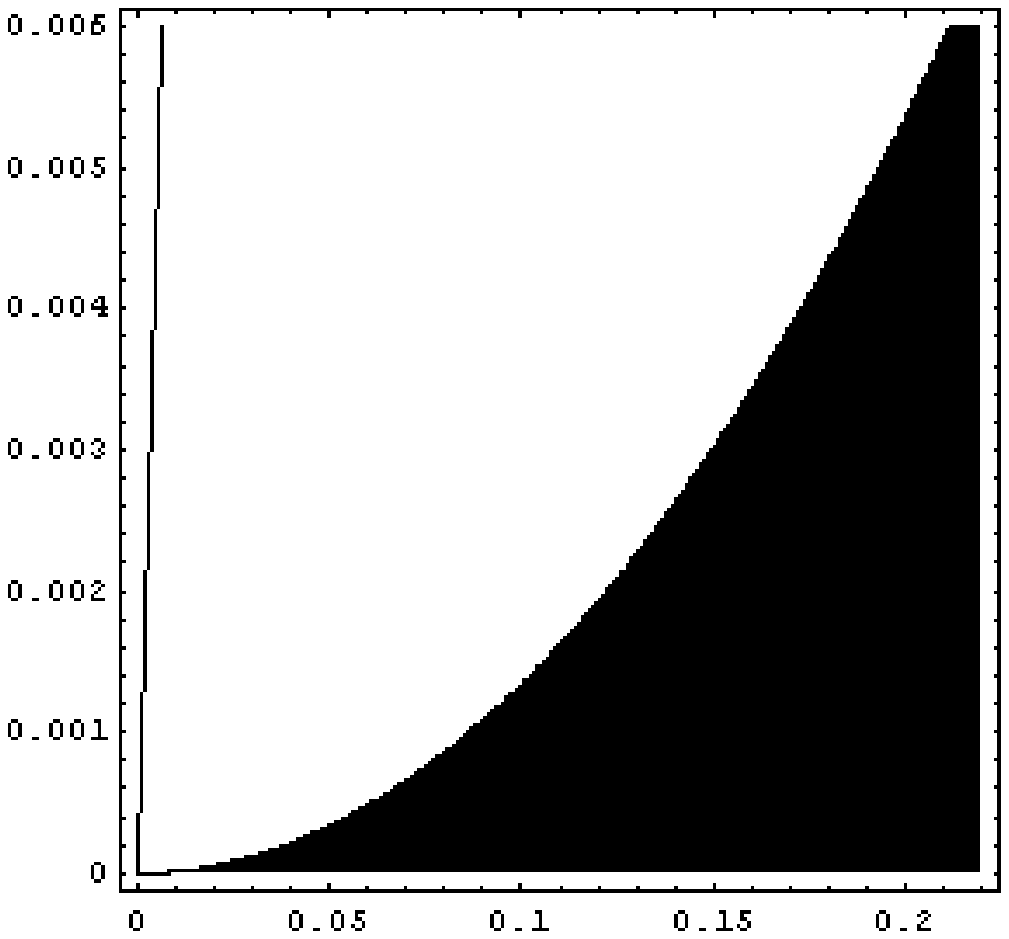}
    \newline
    (c)
    \end{center}
   \end{minipage}
 \end{minipage} 
 }
 \caption{\small{Figs.~(a-b) plot the $A_{sl}$ asymmetry in different ranges of $\Dt$ for different values of $\w$: $\w=0$ (solid-line), $\w=0.001$ (short-dashed), $\w=0.003$ (medium-dashed) and $\w=0.005$ (long-dashed), where the last two are not plotted in Fig.~(b).  Fig.~(c) plots in $|\w|\ vs.\ \Dt(\G^{-1})$ the level curve of Eq.~(2.5), the white area represents the measurable region, which is quite favoured in $\Dt$ with respect to the peak (tiny line in the left).  Different phases $\Omega$ may be visualized in all three plots through the modification in the peak's height in Eq.~(2.4). In all plots we have taken $Re(\epsilon)/1+|\epsilon|^2\approx 0.001$, according to Standard Model expectations.}}
\label{fig1}
\end{figure}

In the $\w=0$ case, the dilepton charge asymmetry it is expected to be {\it exactly} time independent:
\be
 A_{sl} = \left. \frac{I(\ell^+,\ell^+,\Dt)-I(\ell^-,\ell^-,\Dt)}{I(\ell^+,\ell^+,\Dt)+I(\ell^-,\ell^-,\Dt)}\right|_{\w=0} = 4 \re + {\cal O}((Re\ \epsilon)^2).
 \label{ak}
 \ee
On the other hand, if $\w\neq0$, the opposite behaviour of the equal-sign dilepton intensities for small $\Dt$'s produces a cancellation in the denominator of their asymmetry and its behaviour changes drastically.  The time-dependence of $A_{sl}$ is highly altered by a pronounced peak whose position and height are
\bea
\Dt_{peak} = \frac{1}{\G} 1.12 |\w| \qquad \mbox{ and } \qquad A_{sl}(\Dt_{peak}) = 0.77 \cos(\Omega) ,
\label{pirulo}
\eea
respectively. This is seen in Fig.~(\ref{fig1}a), where the $A_{sl}$ asymmetry is plotted for short $\Dt$'s and different values of $\w$.  Since the intensity is quasi-periodic for $\DG/\G \ll 1$, then we expect a second peak-behaviour to be found again around $\Dm\Dt\approx 2\pi$ ($\Dt\approx 8.2 \G^{-1}$), see Fig.~(\ref{fig1}b).
 
Since the main difference between the $\w=0$ and $\w\neq0$ cases is that in the latter the
 asymmetry is time-dependent we may define, with operative purposes, a criterion of
experimental observability regarding the value of the time
derivative $d A_{sl} /d \Dt$.  We may {\it expect} that the limit detectable time furnishes
 \be
 \frac{1}{\G}\left| \frac{d A_{sl} }{d \Dt}(\Dt_{limit} ) \right|= 0.1,
 \label{carlinha}
 \ee 
and hence --in reference to the first peak-- for $\Dt < \Dt_{limit}$ it is (expected to be)
possible to observe the effect of $\w$ as a time-dependence in the $A_{sl}$ asymmetry.  In
Fig.~(\ref{fig1}c) it is plotted the level curve of $\frac{1}{\G}|d A_{sl} /d
\Dt| = 0.1$ as a function of $|\w|$ and $\Dt$ for $\Omega=0$.  As it
can be seen, a value for instance of $\w \sim5\times 10^{-3}$ gives a
$\Dt_{limit} \sim 0.2 \Gamma^{-1}$ as compared to
$\Dt_{peak}=0.005\G^{-1}$.  We conclude that, in much later $\Dt$'s
than $\Dt_{peak}$ the time dependence is still detectable.

The experimental measurements of the $A_{sl}$ asymmetry \cite{exp}, unfortunately, have been performed in the $0.8 \G^{-1} \lesssim \Dt \lesssim 10 \G^{-1}$ region, and hence have not explored effectively the interesting regions.  The short-$\Dt$-peak and its tail are out of the explored region, and the region of the second peak is poor in statistics, since the asymmetry at those times is suppressed by a $e^{-2\pi\G/\Dm}\sim 10^{-4}$ factor.  

Although the experimentally explored region is not the optimal to see the effects of $\w$, we can still put {\it indirect} limits on it by using the available fitted-to-a-constant data of the asymmetry, $A_{sl}^{exp}$.  By requiring that $A_{sl}(\Dt,\w)$ lies, when statistically weighted, within the two standard-deviations of $A_{sl}^{exp}$, we find the $95\%$ confidence limit allowed values for $\w$, namely
 \bea
 -0.0084 \leq Re(\w) \leq 0.0100 \qquad ~ \quad 95\%\mbox{C.L.}
 \eea
These are the first known limits on $\w$.

In order to inquire deeper the possible existence of CPT violation through indistinguishability of particle-antiparticle, the results of this work endorse the experimentalists to either explore the small-$\Dt$ region to look for possible effects of the first peak, or either accumulate statistics to study the region of the second peak with a higher $\Dt$-precision.
\vskip .3cm
We are grateful to J.~Papavassiliou and N.~Mavromatos for useful discussions.



\begin{thebibliography}{99}

\bibitem{prl}
  J.~Bernabeu, N.~E.~Mavromatos and J.~Papavassiliou,
  Phys.\ Rev.\ Lett.\  {\bf 92}, 131601 (2004).

\bibitem{plb}
  E.~Alvarez, J.~Bernabeu, N.~E.~Mavromatos, M.~Nebot and J.~Papavassiliou,
  Phys.\ Lett.\ B {\bf 607}, 197 (2005).

\bibitem{next}
  E.~\'Alvarez, J.~Bernab\'eu and M.~Nebot, to be sent to Physics Letters B.
\bibitem{exp}
  B.~Aubert {\it et al.}  [BABAR Collaboration],
  Phys.\ Rev.\ Lett.\  {\bf 88} (2002) 231801; 

  E.~Nakano {\it et al.}  [Belle Collaboration],
  arXiv:hep-ex/0505017.



\end{thebibliography}
\end{document}